# Fuzzy Logic Trajectory Tracking Controller for a Tanker

DUR MUHAMMAD PATHAN*, MUKHTIAR ALI UNAR**, AND ZEESHAN ALI MEMON*

## ABSTRACT

This paper proposes a fuzzy logic controller for design of autopilot of a ship. Triangular membership functions have been use for fuzzification and the centroid method for defuzzification. A nonlinear mathematical model of an oil tanker has been considered whose parameters vary with the depth of water.

The performance of proposed controller has been tested under both course changing and trajectory keeping mode of operations. It has been demonstrated that the performance is robust in shallow as well as deep waters.

## 1. INTRODUCTION

Design of ship control system has been great challenge since long time, because the dynamics of ship are not only nonlinear but change with operating conditions (depth of water and speed of vehicle). These are also highly influenced by unpredictable external environmental disturbances like winds, sea currents and wind generated waves. IFAC (International Federation of Automatic Control) has identified the ship control system as one of the benchmark problems, which are difficult to solve [1].

In 1911 Elmer Sperry constructed first automatic steering mechanism by developing a closed loop system for a ship [2-3]. Minorsky [4] extended the work of Sperry by giving detailed analysis of position feedback control and formulated three term control law which is referred to as PID (Proportional Integral Derivative) control. Until the 1960s this type of controller was extensively used but after that other linear autopilots like LQG and $H_\infty$ have been reported [5-7].

Nonlinear control schemes such as state feedback linearizaton, backstepping, output feedback control, Lyapunov methods and sliding mode control [8-12] have also been proposed. Applications of nonlinear control techniques depend on exact knowledge of plant dynamics to be controlled. Since the marine vehicle dynamics are highly nonlinear and are coupled with hydrodynamics therefore it is difficult to obtain exact dynamic model of a marine vehicle. To overcome these difficulties, the model free control strategies like Fuzzy Logic and Neural Networks are proved considerably useful.

Yang, et. al. [13-14] discussed the application of fuzzy control for nonlinear systems and applied Takagi-Sugeno type autopilot for a ship to maintain its heading. Santos, et. al. [15] proposed the fuzzy autopilot along with PID controller for control of vertical acceleration for fast moving ferries. Velagic, et. al. [16] developed ship autopilot for track keeping by using Sugeno type fuzzy system.

\* Assistant Professor, Department of Mechanical Engineering, Mehran University of Engineering & Technology, Jamshoro.
\*\* Professor, Department of Computer Systems Engineering, Mehran University of Engineering & Technology, Jamshoro.

Gerasimos, et. al. [17] worked on adaptive Fuzzy control for ship steering. Seo, et. al. [18] worked for ship steering control by using ontology-based fuzzy support agent. Yingbing, et. al. [19] used fuzzy logic for improving the performance of PID controller. Yanxiang, et. al. [20] developed ship steering autopilot by combing the neural network and fuzzy logic techniques. Xiaoyun, et. al. [21] have discussed the advantages of using fuzzy controller along with PID controller for track keeping ship control system. Guo, et. al. [22] presented combination of neural and fuzzy controller for modeling and control of ship.

The above mentioned work is based on the integration of fuzzy logic with other techniques. Moreover, none of the authors have taken the depth of water into consideration. In this work a fuzzy logic controller is proposed which yields robust performance at different depths of sea water.

## 2. DYNAMICS OF SHIP

Marine vehicles consist of submersible and surface vehicles. The maneuverability of marine vehicles is completely described by their dynamics; dynamics of marine vehicles consist of kinematics and kinetics. The study of kinematics provides the complete description of motion of vehicles. It develops relation between the body fixed linear and angular velocities and earth fixed linear and angular velocities. The kinetics studies dynamic response when forces and moments are applied to it. It studies the rigid body and hydrodynamic forces and moments acting on the body.

Motion of marine vehicles is described as six degree of freedom of motion because six independent coordinates (surge, sway, heave, yaw, roll and pitch) are required to specify the complete motion of vehicles as given in Table 1, where the first three co-ordinates and their time derivatives describe the position and translational motion along x-, y-, and z-axis. The last three coordinates and their time derivatives correspond to orientation and rotational motion of the vehicle.

### 2.1 Heading Motion of Ship

The motion of ship is regarded as surface motion and its motion is confined in course of earth fixed $X_E$-$Y_E$ plane instead of space, hence the heading motion of ship is considered to be three degree of freedom motion because only three independent coordinates (sway, yaw and psi) are required to specify the heading motion of vehicle as shown in Fig. 1.

### 2.2 The ESSO 190000 dwt Tanker

The ESSO is a large size tanker of 304.8m length. The mathematical model describing the maneuverability of this tanker is given in [23]. The parameters of ESSO are listed below [7]:

Length between perpendiculars = 304.8m  
Beam = 47.17m  
Draft to design waterline = 18.46m  
Displacement = 220000 m$^3$  
Block coefficient = 0.83  
Design speed = 16 knots  
Nominal Propeller = 80 rpm  

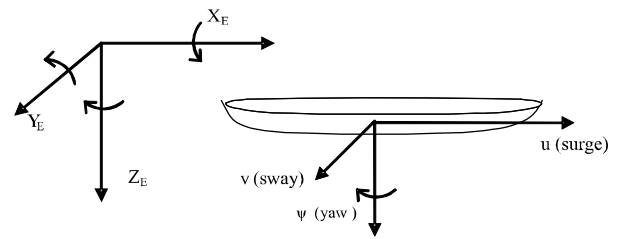

*FIG. 1. SHIP FIXED AND EARTH FIXED AXES*

**TABLE 1. COMPONENTS OF MARINE VEHICLE**

| DOF | Motion Components | Forces and Moments | Linear and Angular Velocities | Position and Angles |
|---|---|---|---|---|
| 1 | Motion in x-Direction (surge) | X | u | x |
| 2 | Motion in y-Direction (sway) | Y | v | y |
| 3 | Motion in z-Direction (heave) | Z | w | z |
| 4 | Rotation about x-axis (roll) | K | p | $\phi$ |
| 5 | Rotation about y-axis (pitch) | M | q | $\theta$ |
| 6 | Rotation about z-axis (yaw) | N | r | $\psi$ |

The model of the tanker is represented as:

$$\underline{\dot{x}} = A\underline{x} + Bu \qquad (1)$$

where x=[u v r] is the state vector of the ship, A is the system matrix, B is the input matrix and $u$=[δr] is input component i.e rudder deflection. The purpose of this investigation is to develop a fuzzy logic control system that yields satisfactory performance for any reference heading angles from ±3° to +45° and also provides satisfactory performance for a predefined trajectory by providing series of heading angles. The performance index is that the ship should track the desired heading with reasonable accuracy. The difference between the desired heading and actual heading (i.e the heading error) should not exceed ±3° and the control signal (rudder angle) should not reach the maximum value of ±35°.

## 3. DEVELOMENT OF FUZZY LOGIC CONTROLLER

The central idea of fuzzy logic control methodology is to map the input states for a particular output. This can be done by using FIS (Fuzzy Inference System). FIS is a tool to formulate the mapping from a given input to specific output using fuzzy logic. This mapping provides the basis from which decisions are made.

Fuzzy controller consists of an input, processing and output stages. The input stage contains membership functions, where the crisp input is mapped to an appropriate membership function and truth value is assigned to crisp input. The processing stage consists of set of rules. This stage invokes each appropriate rule and generates results for each rule, then combines the results of the active rules. The output stage consists of defuzzifacation function, where combined result of the fuzzy rules is converted into crisp output.

Generally the development of control system consists of the following steps:

- Selection of appropriate inputs and output variables for control system.
- Selection of appropriate ranges for inputs and output variables.
- Selection of appropriate membership function for each fuzzy input and output.
- Development of subsets of input and output variables and labeling each of subset into linguistic (fuzzy) variables
- Development of appropriate rules and processing of rules.
- Defuzzification of output of the system into crisp output.

### 3.1 Inputs and Outputs for Control System

The component which affects the heading motion of ship is known as rudder. The states which are significantly affected by rudder deflection (δr) are heading/yaw angle (ψ), heading rate/yaw rate (r) and sway velocity (v). The performance of the controller is based on minimization of the difference between desired and actual states (i.e error). The desired states are generated from desired model by using second order differential equation. Hence inputs of the fuzzy controller are heading angle error (ψ_error) and heading rate error (r_error). The output of controller is chosen as rudder deflection (δr).

### 3.2 Ranges for Input and Output

In ranges for variables are selected as; from -3° to +3° for heading error and -35° to +35° for rudder deflection. Starting from these limits, the values for ranges of input and output variables are obtained by trial and error (Table 2). Ranges are selected from negative to positive so that ship may attain clockwise or anticlockwise direction (from port to starboard).

Fuzzy Logic Trajectory Tracking Controller for a Tanker

## 3.3 Membership Function

Membership function is a function which assigns the input a value between 0 and 1. It exhibits degree of membership of a member in a particular fuzzy set. In this work triangular function is used for fuzzification, which gives smooth transition from initial to final state. A function theoretic form of triangular function is given by Equation (2). The Equation (2) is derived in such a manner that it assigns any input a value between 0-1.

$$\mu(x_{error}, a, b, c) = \begin{cases} 0 & x_{error} < a, x_{error} > c \\ \dfrac{x_{error} - a}{b - a} & a \leq x_{error} \geq b \\ 1 & x_{error} = b \\ \dfrac{c - x_{error}}{c - b} & b \leq x_{error} \geq c \end{cases} \quad (2)$$

where $x_{error}$ are crisp values of input state errors, 'μ' is membership value assigned to input, 'a' and 'c' describe the limits of function, 'b' is core (having membership of one) of triangular membership. The values of limits and core for different variables are given in Tables 3-5.

**TABLE 2. UNIVERSAL SETS FOR RANGES OF INPUT AND OUTPUT VARIABLES**

| Input Vector ($x_{error}$) | heading error | ψ_error | -0.4 to -0.4 deg. |
|---|---|---|---|
| | heading rate error | r_error | -0.01 to -0.01 deg./second |
| Output (u) | rudder deflection | δr | -0.8 to -0.8° deg. |

**TABLE 3. FUZZY VARIABLES AND SUBSETS OF HEADING_ERROR**

| Fuzzy Subsets | Heading error (ψ_error) | | |
|---|---|---|---|
| | a | b | c |
| BN | -0.530 | -0.400 | -0.266 |
| MN | -0.400 | -0.266 | -0.133 |
| SN | -0.266 | -0.133 | 0.000 |
| ZE | -0.133 | 0.000 | 0.133 |
| SP | 0.000 | 0.133 | 0.2660 |
| MP | 0.133 | 0.266 | 0.400 |
| BP | 0.266 | 0.400 | 0.530 |

## 3.4 Subsets of Ranges and Fuzzy Variables

Each of the universal sets of inputs (ψ_error and r_error) and ouput (δr) are divided into seven subsets. Each subset is labeled by linguistically defined fuzzy variables: BN (Big Negative), MN (Medium Negative), SN (Small Negative), ZE (Zero), SP (Small Positive), MP (Medium Positive) and BP (Big Positive). The ranges of subsets are defined is such a manner that their values overlap each other as given in Tables 3-5.

Equation (2) is used to map each of these subsets to a triangular membership function as shown in Fig. 2, where the crisp input is assigned a value between 0 and 1. It exhibits degree of membership of particular input to particular fuzzy sets. As the value of variable changes, its membership to particular fuzzy subset varies from 0-1 and then from 1-0. Because of overlapping the particular value of the subset can be member of more than one fuzzy subset. As the value of variable changes the membership value tracks from one fuzzy set to another. This interpolation between the fuzzy sets helps to make decisions.

**TABLE 4. FUZZY VARIABLES AND SUBSETS OF HEADING_RATE_ERROR**

| Fuzzy Subsets | Heading rate error (r_error) | | |
|---|---|---|---|
| | a | b | c |
| BN | -0.133300 | -0.010000 | -0.006665 |
| MN | -0.010000 | -0.006665 | -0.003335 |
| SN | -0.006665 | -0.003330 | 0.000000 |
| ZE | -0.003335 | 0.000000 | 0.003335 |
| SP | 0.000000 | 0.003335 | 0.006665 |
| MP | 0.003335 | 0.006665 | 0.010000 |
| BP | 0.006665 | 0.010000 | 0.133300 |

**TABLE 5. FUZZY VARIABLES AND SUBSETS OF RUDDER DEFLECTION**

| Fuzzy Subsets | Rudder deflection (δr) | | |
|---|---|---|---|
| | a | b | C |
| BN | -1.0670 | -0.8000 | -0.5333 |
| MN | -0.8000 | -0.5333 | -0.2667 |
| SN | -0.5333 | -0.2667 | 0.0000 |
| ZE | -0.2667 | 0.0000 | 0.2667 |
| SP | 0.0000 | 0.2667 | 0.5333 |
| MP | 0.2667 | 0.5333 | 0.8000 |
| BP | 0.5333 | 0.8000 | 1.0670 |

## 3.5 Rules and Processing of Rules

The processing stage of fuzzy controller is the collection of rules. These rules are written in the form of IF THEN statements. IF part of the rule is called Antecedent, while THEN part is called Consequent. If there are more than one inputs or outputs, then they are combined by fuzzy logical operators generally AND (min) or OR (max). The number of rules depends upon the number of inputs, number of subset/membership functions and number of output.

In this work the controller consists of two inputs and one output. Each input and output variable is divided into seven subsets. It is observed that the square number of rules (Table 6) helps to maintain the structural symmetry of the system. For the proposed controller, 7x7=49 rules are developed. Some of the rules are given below:

Rule 1: IF r_error is big negative AND ψ_error is big negative THEN δr is big negative

Rule 2: IF r_error is medium negative AND ψ_error is big negative THEN δr is big negative

Rule 3: IF r_error is small negative AND ψ_error is big negative THEN δr is medium negative

Rule 4: IF r_error is zero AND ψ_error is big negative THEN δr is medium negative

Rule 5: IF r_error is small positive AND ψ_error is big negative THEN δr is small negative

Rule 6: IF r_error is medium positive AND ψ_error is big negative THEN δr is small negative.

.
.
.

and so on

A summary of complete 49 rules can be written in fuzzy associative matrix form as given in Table 6.

In Table 6, it is observed that the main diagonal of the table consists of zeros and on the either side of this diagonal the fuzzy variables change gradually having the opposite signs. This condition maintains the structural symmetry of the plant under consideration.

**TABLE 6. SUMMARY OF RULES**

| r_error / ψ_error | BN | MN | SN | ZE | SP | MP | BP |
|---|---|---|---|---|---|---|---|
| BN | BN | BN | MN | MN | SN | SN | ZE |
| MN | BN | MN | MN | SN | SN | ZE | SP |
| SN | MN | MN | SN | SN | ZE | SP | SP |
| ZE | MN | SN | SN | ZE | SP | SP | MP |
| SP | SN | SN | ZE | SP | SP | MP | MP |
| MP | SN | ZE | SP | SP | MP | MP | BP |
| BP | ZE | SP | SP | MP | MP | BP | BP |

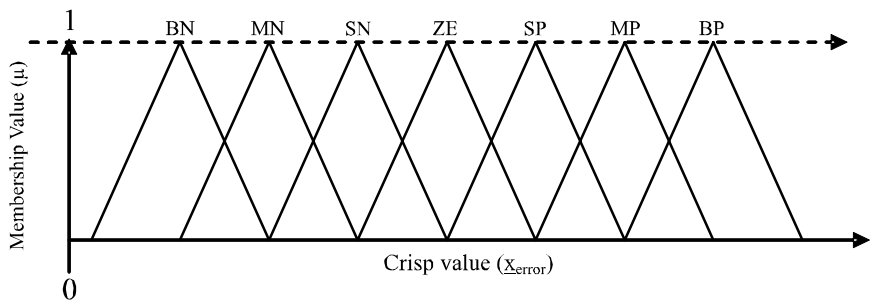

*FIG. 2. FUZZIFICATION BY USING TRIANGULAR MEMBERSHIP FUNCTION*

## 3.6 Defuzzification

Defuzzification is a process of converting the aggregated output of the rule into a single crisp value. Various methods of defuzzificaion are available. In this work centroid method is used. In this method the centre of area under the curve is calculated by using Equation (3).

$$u = \frac{\int \mu(z).z.dz}{\int \mu(z)dz} \qquad (3)$$

where 'z' is aggregated output of active rules and 'u' is crisp output of fuzzy controller.

## 4. SIMULATION OF FUZZY CONTROL SYSTEM

In this section simulation results are presented. The simulations are carried by developing a closed loop system consisting of desired system model, fuzzy controller and model of the ship as shown in Fig. 3. From desired model desired states are generated by step command input, however the actual states generated by ship model are fed back to comparison element. From where the error signals are generated which are fed to the fuzzy controller to minimize the error and regulate the ship in the desired direction. Simulations are carried out in MATLAB Version 7.

In simulation studies test conditions are set by changing the depth of water to observe the effectiveness of the controller.

For model of the tanker, deep and confined waters are described by a parameter called depth ratio ($\zeta$). This ratio is calculated by Equation (4) [7]:

$$\zeta = \frac{T}{h-T} \qquad (4)$$

where h is water depth and T is draft to design waterline. Water depth should be greater than the draft. The relation between $\zeta$ and h provides a transition point where hydrodynamic coefficient $Y_{uv\zeta}$ changes value and it obeys the following condition:

if $\zeta$ is less than 0.8, then $Y_{uv\zeta} = 0$

if $\zeta$ is greater than or equal to 0.8, then $Y_{uv\zeta} = -0.85(1-0.8/\zeta)$ [7,19-20].

By analyzing this relation it is observed that at depth greater than 100m, the depth ratio $\zeta$ does not vary considerably. Therefore two depth regions (i.e greater and less then 100m) are normally considered for test conditions, where the dynamics of the vehicle changes.

In this work, the simulations are carried out for various command angles in shallow water at 24m depth and deep water at 200m depth of water.

## 5. RESULTS AND DISCUSSION

Various simulation are carried out by using the closed loop system, the graphical results are presented in Figs. 4-6.

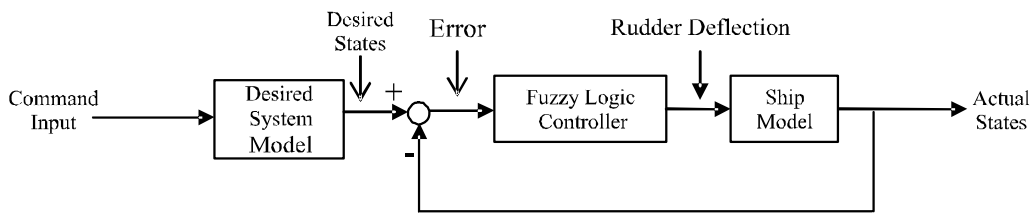

*FIG. 3. FUZZY CONTROL SYSTEM*

### 5.1 Simulation for Heading Control at Shallow and Deep Waters

Results presented in Figs. 4-5 show the performance of heading controller for 45° command angles. The simulations are carried out for shallow and deep waters. In Figs. 4(a)-5(a), the dashed lines show the desired heading and solid lines represent the actual output of the controller. Figs. 4(b)-5(b) represent the difference between the desired and actual heading (heading error). Figs. 4(c)-5(c) show the behaviour of the rudder in response to command angles.

Fig. 5 evaluates the performance of controller at 45° command heading angles in shallow waters (24m depth). In Fig. 5 it is observed that the actual response follows the desired response. The maximum errors calculated for 45° command heading angles is -2.8°.

However, Fig. 5 exhibits the performance of the controller for same command angle in deep waters (200m depth). In Fig. 5 it is observed that the actual response follows the desired response with some oscillations on either side of the actual response. However, maximum values of errors remain within the specified limits.

### 5.2 Simulation for Trajectory Tracking Control at Continuous Varying Depth of Water

Fig. 6 shows the results of the controller for trajectory tracking by providing a series of 10, 20 and -5° of heading commands at continuous varying depths. Fig. 6(a) represents the desired trajectory by dashed lines and actual trajectory by solid lines. Fig. 6(b) represents the difference between the desired and actual trajectory (error). Fig. 6(c) shows the behaviour of the rudder in response to series of command angles. Fig. 6(d) represents continuously varying depths ranging from 24-200m.

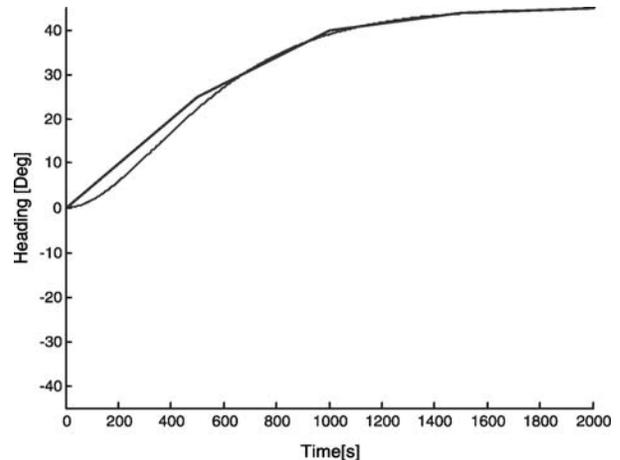

*(a) DESIRED RESPONSE AND ACTUAL RESPONSE FOR 45° COMMAND ANGLES*

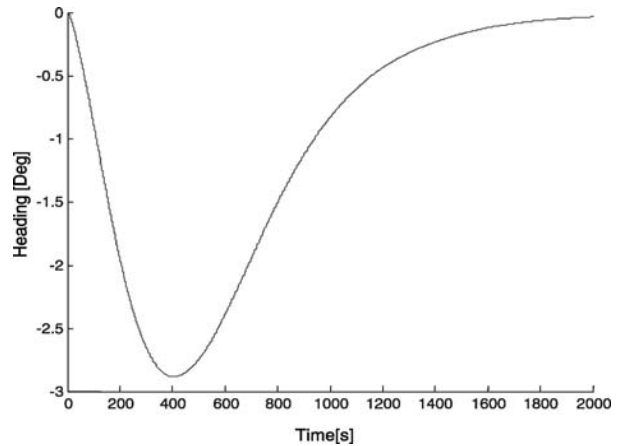

*(b) DIFFERENCE BETWEEN DESIRED AND ACTUAL HEADING (I.E HEADING ERROR)*

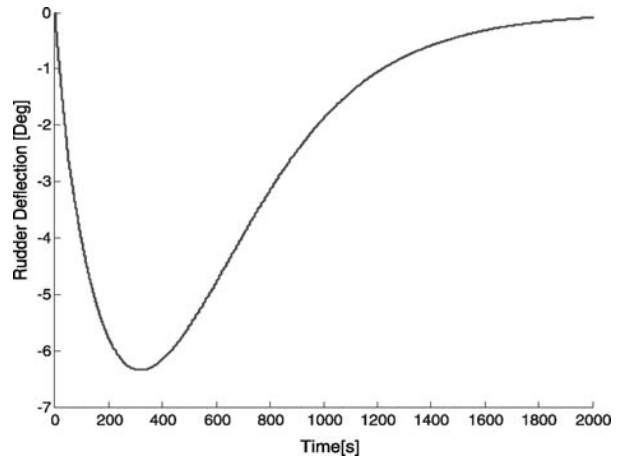

*(c) CONTROL COMPONENT EFFORT (RUDDER DEFLECTION)*

FIG. 4. SIMULATION RESULTS AT 45° COMMAND HEADING ANGLE AT 24M DEPTH OF WATER

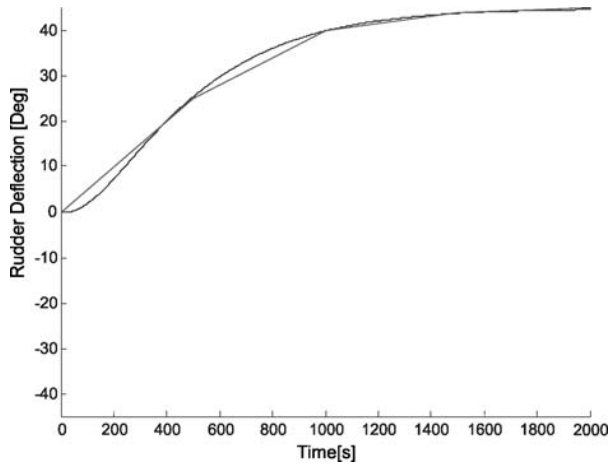

*(a) 45º DESIRED AND ACTUAL HEADING ANGLE*

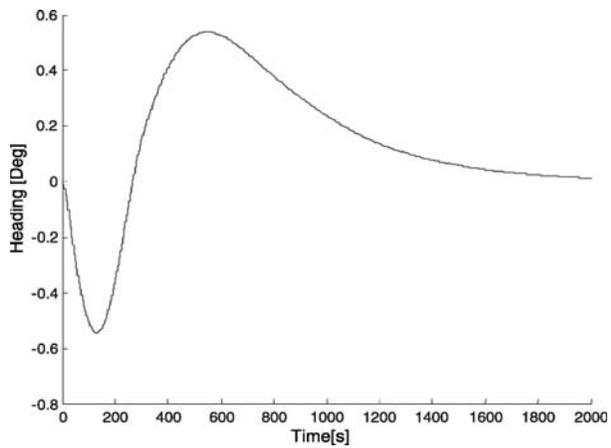

*(b) DIFFERENCE BETWEEN DESIRED AND ACTUAL HEADING (I.E HEADING ERROR)*

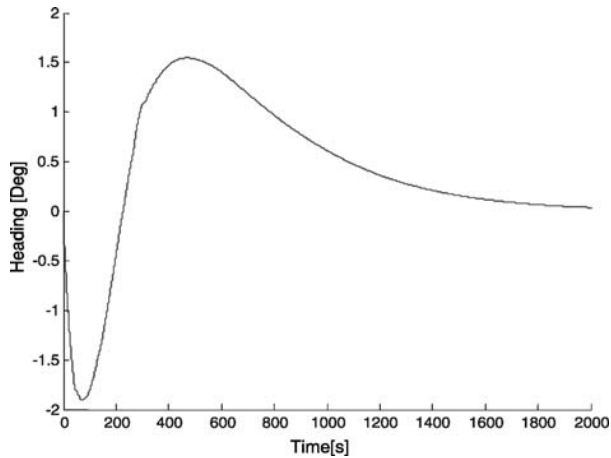

*(c) CONTROL COMPONENT EFFORT (RUDDER DEFLECTION)*

FIG. 5. SIMULATION RESULTS AT 45º COMMAND HEADING ANGLE AT 200 M DEPTH OF WATER

## 5.3 Observation and Result Discussion

From Figs. 4(a)-5(a), it is observed that the tanker takes approximately 2000 seconds to reach the steady state position. This is because of the large size of the vehicle.

Results show that in shallow water the actual response tracks the desired response smoothly; however in deep sea water the actual response oscillates around the desired response, this is because of variation in dynamics of ship due to change in depth of water. Despite the change of dynamics of the ship, the performance of the fuzzy controller remains robust and response converges to steady state condition within a limited time. Similarly for trajectory tracking the error remains within the specified limits.

## 6. CONCLUSION

In this paper a Fuzzy logic controller is designed for heading and trajectory tracking control of a tanker. The development of the controller is based on Mamdani type FIS. The controller is designed in such a manner that the tanker can turn on both sides starboard and port of the vehicle. The performance of the controller is tested in shallow waters and deep sea waters. It is concluded that the overall performance of the controller is satisfactory both in course changing and course keeping. The error in all the cases remains within the specified limits and controller has capability to cope with variations in dynamics of the system.

## ACKNOWLEDGEMENTS

The first author acknowledges the Higher Education Commission of Pakistan for awarding funds to conduct this research at Mehran University of Engineering & Technology, Jamshoro, Pakistan.

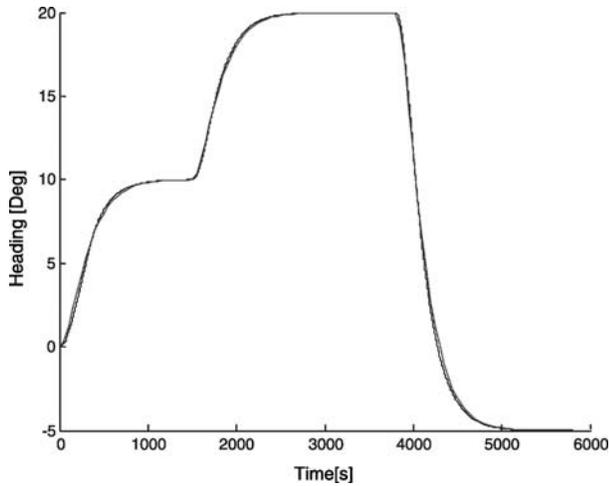

*(a) TRAJECTORY TRACKING A SERIES OF COMMAND ANGLES (IE 10º, 20º AND -5º)*

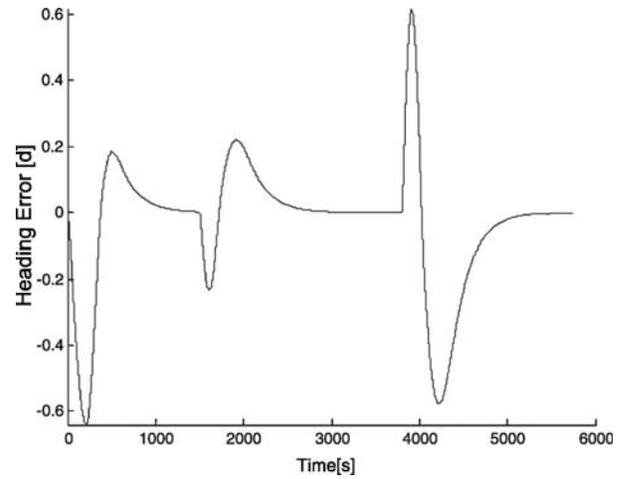

*(b) DIFFERENCE BETWEEN DESIRED AND ACTUAL HEADING (IE HEADING ERROR)*

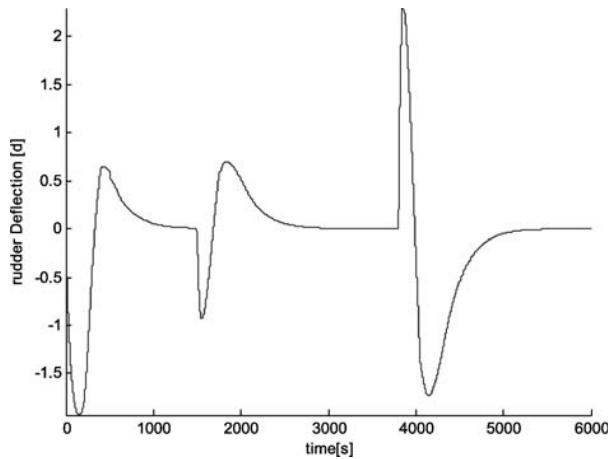

*(c) CONTROL COMPONENT EFFORT (RUDDER DEFLECTION)*

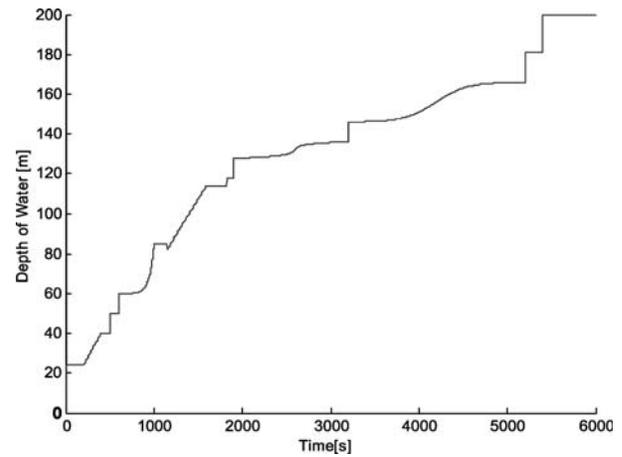

*(d) VARYING DEPTH OF WATER FROM 24-200M DEPTH*

FIG. 6. RESULTS OF CONTROLLER FOR TRAJECTORY TRACKING AT CONTINUOUS VARYING DEPTH FROM 24-200 M.